\documentclass[letterpaper,twocolumn]{jpsj3}
\usepackage{txfonts}
\usepackage{multicol}
\usepackage{nidanfloat}
\usepackage{color}

\def\gsim{\mathop {\vtop {\ialign {##\crcr 
$\hfil \displaystyle {>}\hfil $\crcr \noalign {\kern1pt \nointerlineskip } 
$\,\sim$ \crcr \noalign {\kern1pt}}}}\limits}
\def\lsim{\mathop {\vtop {\ialign {##\crcr 
$\hfil \displaystyle {<}\hfil $\crcr \noalign {\kern1pt \nointerlineskip } 
$\,\,\sim$ \crcr \noalign {\kern1pt}}}}\limits}

\title{Gr\"{u}neisen Parameter and Thermal Expansion by the Self-Consistent Renormalization Theory of Spin Fluctuations}

\author{Shinji Watanabe$^1$ and Kazumasa Miyake$^2$}
\inst{$^1$Department of Basic Sciences, Kyushu Institute of Technology, Kitakyushu, Fukuoka 804-8550, Japan \\
$^2$
Center for Advanced High Magnetic Field Science, Osaka University, Toyonaka 560-0043, Japan
}

\abst{
The thermal expansion coefficient $\alpha$ and the Gr\"{u}neisen parameter $\Gamma$ near the magnetic quantum critical point (QCP) are derived on the basis of the 
self-consistent renormalization (SCR) theory of spin fluctuation. 
From the SCR entropy, the specific heat $C_{V}$, $\alpha$, and $\Gamma$ are shown to be expressed in a simple form as $C_{V}=C_{\rm a}-C_{\rm b}$, $\alpha=\alpha_{\rm a}+\alpha_{\rm b}$, and $\Gamma=\Gamma_{\rm a}+\Gamma_{\rm b}$, respectively, where $C_{\rm i}$, $\alpha_{\rm i}$, and $\Gamma_{\rm i}$ $({\rm i}={\rm a}, {\rm b})$ are related with each other. As the temperature $T$ decreases, $C_{\rm a}$, $\alpha_{\rm b}$, and $\Gamma_{\rm b}$ become dominant in $C_{V}$, $\alpha$, and $\Gamma$, respectively. The inverse susceptibility of spin fluctuation coupled to the volume $V$ in $\Gamma_{\rm b}$ is found to give rise to the divergence of $\Gamma$ at the QCP for each class of ferromagnetism and antiferromagnetism (AFM) in spatial dimensions $d=3$ and $2$. This $V$-dependent inverse susceptibility in $\alpha_{b}$ and $\Gamma_{\rm b}$ contributes to the $T$ dependences of $\alpha$ and $\Gamma$, and even affects their criticality in the case of the AFM QCP in $d=2$. $\Gamma_{\rm a}$ is expressed as $\Gamma_{\rm a}(T=0)=-\frac{V}{T_0}\left(\frac{\partial T_0}{\partial V}\right)_{T=0}$ with $T_0$ being the characteristic temperature of spin fluctuation, which has an enhanced value in heavy electron systems. 
}

\begin{document}
\maketitle

\section{Introduction}

Quantum critical phenomena in itinerant electron systems have attracted considerable attention in condensed matter physics. 
When the continuous transition temperature of the magnetically ordered phase is suppressed to absolute zero 
by changing control parameters such as pressure and magnetic field,  
the quantum critical point (QCP) is realized. 
Near the QCP, the enhanced spin fluctuation causes the non-Fermi liquid behavior in physical quantities, which is called quantum critical phenomenon.

The self-consistent renormalization (SCR) theory of spin fluctuations developed by Moriya and coworkers has succeeded in explaining not only the Curie--Weiss behavior   
but also the quantum critical behavior in the magnetic susceptibility in the case of ferromagnetic criticality~\cite{Moriya}. 
The SCR theory has also explained the quantum criticality in other physical quantities such as resistivity, specific heat, and NMR relaxation rate in the ferromagnetic and antiferromagnetic cases~\cite{HM1995,MT,IM1996,IM1998,Ishigaki}, which has been endorsed by the 
renormalization group (RG) analysis by Hertz~\cite{Hertz} and Millis~\cite{Millis}. 

The magneto-volume effect in nearly ferromagnetic (FM) metals~\cite{MU1980} and in antiferromagnetic (AFM) metals~\cite{IM1998} has been discussed by Moriya and coworkers. 
Kambe et al. pointed out a possibility that the Gr\"{u}neisen parameter $\Gamma$~\cite{Gruneisen1912} diverges at the QCP~\cite{Kambe1997}.  
By using the scaling hypothesis and the RG theory, 
Zhu et al. evaluated the critical part of the thermal expansion coefficient and the Gr\"{u}neisen parameter, and showed that $\Gamma$ diverges at the QCP~\cite{Zhu2003,Garst2005}. 

In this paper, we derive   
the thermal expansion coefficient and the Gr\"{u}neisen parameter near the magnetic QCP 
on the basis of the SCR theory. 
By using the Maxwell relation, we show that $\alpha(T)$ derived from the pressure derivative of the entropy has a much simpler form than that derived from the temperature derivative of the pressure by the spin fluctuation theory~\cite{Takahashi2006}. Our result makes it possible to clarify the origin of the divergence of $\Gamma$ at the QCP, explicitly showing  
 that the inverse susceptibility of spin fluctuation coupled to the volume gives rise to the divergence of $\Gamma$ at the QCP. 
Numerical calculations of $\alpha(T)$ and $\Gamma(T)$ are also performed for the FM QCP and AFM QCP in three- and two-spatial dimensions, which reveal that the quantum critical behavior appears at a sufficiently lower temperature than the characteristic temperature of spin fluctuation. 
We find that the volume derivative of the mode-mode coupling of spin fluctuation contributes to the temperature dependences of $\alpha(T)$ and $\Gamma(T)$ and even affects the quantum criticality at the AFM QCP in the two--spatial dimension.

This paper is organized as follows. 
In Sect.~2, the SCR theory is outlined and the nature of the specific heat near the QCP is summarized. 
The thermal expansion coefficient and the Gr\"{u}neisen parameter are analyzed in Sects.~3 and 4, respectively. 
Comparison with experiments is discussed in Sect.~5. 
The paper is summarized in Sect.~6.  

\section{SCR Theory}

The SCR theory of spin fluctuations is outlined. 
First, the formalism of the SCR theory is explained in Sect.~2.1. 
In Sect.~2.2, the critical properties of the specific heat derived from the entropy are summarized. 
Hereafter, the energy units are taken as $\hbar=1$ and $k_{\rm B}=1$. 

\subsection{Formalism of the SCR theory}

The action of interacting electrons is given in the form of the Ginzburg--Landau--Wilson functional 
\begin{eqnarray}
\Phi[\varphi]=\frac{1}{2}\sum_{\bar{q}}\Omega_{2}(\bar{q})\varphi(\bar{q})\varphi(-\bar{q}) 
\nonumber
\\
+\sum_{\bar{q}_1,\bar{q}_2,\bar{q}_3,\bar{q}_4}
\Omega_{4}(\bar{q}_1, \bar{q}_2, \bar{q}_3, \bar{q}_4)
\varphi(\bar{q}_1)\varphi(\bar{q}_2)\varphi(\bar{q}_3)\varphi(\bar{q}_4)
\delta\left(\sum_{i=1}^{4}\bar{q}_i\right), 
\label{eq:Action}
\end{eqnarray}
which has been derived from the Hamiltonian via the Hubbard--Stratonovich transformation applied to the onsite Coulomb interaction~\cite{Hertz}. 
Here, $\bar{q}$ is the abbreviation for $\bar{q}\equiv({\bf q}, {\rm i}\omega_{l})$, where  $\omega_l=2\pi lT$ $(l=0, \pm1, \pm2, \cdots)$ is the Matsubara frequency with $T$ being the temperature. 
Note that Eq.~(\ref{eq:Action}) has the form derived from a single component of quadratic spin interaction (e.g., $S_i^{z}S_i^{z}$)~\cite{Hertz}.  
In the case of the isotropic Heisenberg interaction (e.g., ${\bf S}_i\cdot{\bf S}_i$), the factor 3 is to be multiplied to the right--hand side of Eq.~(\ref{eq:Action})~\cite{Ishigaki,Takahashi1999,note_Heisenberg}. 
In critical phenomena, 
long wavelength $|{\bf q}|\ll q_{\rm c}$ around the magnetically ordered vector ${\bf Q}$ and the low--frequency $|\omega|\ll\omega_{\rm c}$ regions play dominant roles with $q_{\rm c}$ and $\omega_{\rm c}$ being the cutoffs for the momentum and frequency, respectively. 
Hence, the coefficients 
$\Omega_i$ for $i=2, 4$ in Eq.~(\ref{eq:Action}) are expanded for $q$ and $\omega$ around $({\bf Q}, 0)$: 
\begin{eqnarray}
\Omega_2({\bf q},{\rm i}\omega_l)\approx\frac{\eta_0+Aq^2+C_{q}|\omega_l|}{N_{\rm F}}, 
\end{eqnarray}
where $C_q$ is defined by $C_{q}\equiv C/q^{z-2}$ with $z$ being the dynamical exponent (e.g., $z=3$ for FM and $z=2$ for AFM) and $N_{\rm F}$ is the density of states at the Fermi level, 
and $\Omega_{4}(\bar{q}_1, \bar{q}_2, \bar{q}_3, \bar{q}_4)\approx v_{4}/(\beta N)$ with $\beta\equiv1/T$.

To construct the action for the best Gaussian taking into account the mode-mode coupling of spin fluctuations up to the fourth order in $\Phi[\varphi]$, we use Feynman's inequality~\cite{Feynman} on the free energy:
\begin{eqnarray}
F\le F_{\rm eff}+T\langle\Phi-\Phi_{\rm eff}\rangle_{\rm eff}\equiv\tilde{F}(\eta). 
\label{eq:Free_SCR} 
\end{eqnarray}
Here, the effective action $\Phi_{\rm eff}$ is given by  
\begin{eqnarray}
\Phi_{\rm eff}[\varphi]=\frac{1}{2}\sum_{l}\sum_{q}\frac{\eta+Aq^2+C_q|\omega_l|}{N_{\rm F}}\left|\varphi(q,{\rm i}\omega_{l})\right|^{2},    
\label{eq:Action_eff}
\end{eqnarray}
where $\eta$ includes the effect of the mode-mode coupling of spin fluctuations and parameterizes the closeness to the quantum criticality. 
By the optimal condition $d\tilde{F}(\eta)/d\eta=0$, the self-consistent equation for $\eta$, i.e., the SCR equation is obtained. 
By introducing the scaled form as $y\equiv\eta/(Aq_{\rm B}^2)$, $x\equiv q/q_{\rm B}$, $x_{\rm c}\equiv q_{\rm c}/q_{\rm B}$, and $t\equiv T/T_0$, where $T_0$ is the characteristic temperature of spin fluctuation defined by 
\begin{eqnarray}
T_0\equiv \frac{Aq_{\rm B}^2}{2\pi C_{q_{\rm B}}},
\label{eq:T0}
\end{eqnarray}
and $q_{\rm B}$ is the wave number characterizing the Brillouin zone, 
the SCR equation in the $d$--dimensional system is expressed as   
\begin{eqnarray}
y=y_0+\frac{d}{2}y_1\int_{0}^{x_{\rm c}}dxx^{d+z-3}
\left\{
{\ln}u-\frac{1}{2u}-\psi(u)
\right\} 
\label{eq:SCReq2}
\end{eqnarray}
for $d+z>4$~\cite{MT,HM1995,IM1996}
and 
\begin{eqnarray}
y=y_0+\frac{y_1}{2}\left(
y{\ln}y
+d
\int_{0}^{x_{\rm c}}dxx
\left\{
{\ln}u-\frac{1}{2u}-\psi(u)
\right\}
\right) 
\label{eq:SCReq3}
\end{eqnarray}
for $d+z=4$~\cite{IM1998}.  
Here, $y_0$ and $y_1$ are constants, $u$ is defined as $u\equiv x^{z-2}(y+x^2)/t$, and      
$\psi(u)$ is the digamma function. 
The solutions of Eqs.~(\ref{eq:SCReq2}) and (\ref{eq:SCReq3}) at the QCP can be obtained by inputting $y_0=0$ with $y_1$ and the cutoff $x_{\rm c}$ being set as constant values, e.g., $y_1=1$ and $x_{\rm c}=1$. 
The low--$t$ behavior of $y$ for each class of the FM $(z=3)$ and AFM $(z=2)$ in $d=3$ and 2, respectively,  is listed in the first column of Table~\ref{tb:CaCb}. 
Note that the criticality in each class coincides with that shown by the RG theory~\cite{Millis} including logarithmic corrections in $d=2$. 

\subsection{Entropy and specific heat}

The entropy $S=-\left(\frac{\partial{\tilde{F}}}{\partial{T}}\right)_{V}$ is obtained by differentiating the free energy $\tilde{F}$ with respect to the temperature under a constant volume as~\cite{Takahashi1999}
\begin{eqnarray}
S&=&-Nd\int_{0}^{x_{\rm c}}dxx^{d-1}
\left\{
{\ln}\sqrt{2\pi}-u+\left(u-\frac{1}{2}\right){\ln}u-{\ln}\Gamma(u)
\right\}
\nonumber
\\
&+&Nd\int_{0}^{x_{\rm c}}dxx^{d-1}u
\left\{
{\ln}u-\frac{1}{2u}-\psi(u)
\right\},
\label{eq:S}
\end{eqnarray}
where $\Gamma(u)$ is the Gamma function.

By differentiating the entropy $S$ in Eq.~(\ref{eq:S}) with respect to 
the temperature under a constant volume~\cite{Takahashi1999,Ishigaki}, the specific heat is obtained as 
\begin{eqnarray}
C_V&=&T\left(\frac{\partial{S}}{\partial{T}}\right)_{V}, 
\nonumber
\\
&=&
C_{\rm a}
-C_{\rm b},
\label{eq:CvT}
\end{eqnarray}
where 
$C_{\rm a}$ and $C_{\rm b}$ are given by 
\begin{eqnarray}
C_{\rm a}&=&-Nd\int_{0}^{x_{\rm c}}dxx^{d-1}u^2\left\{
\frac{1}{u}+\frac{1}{2u^2}-\psi'(u)
\right\}, 
\label{eq:Ca}
\\
C_{\rm b}&=&
\tilde{C}_{\rm b}
\left(\frac{\partial{y}}{\partial{t}}\right)_{V},
\label{eq:Cb}
\end{eqnarray}
respectively. Here, $\psi'(u)$ is the trigamma function and $\tilde{C}_{\rm b}$ is given by 
\begin{eqnarray}
\tilde{C}_{\rm b}&=&-Nd\int_{0}^{x_{\rm c}}dxx^{d+z-3}u\left\{
\frac{1}{u}+\frac{1}{2u^2}-\psi'(u)
\right\}. 
\label{eq:tildaCb}
\end{eqnarray}
The derivative $\left(\frac{\partial{y}}{\partial{t}}\right)_{V}$ in Eq.~(\ref{eq:Cb}) can be calculated explicitly, by differentiating the SCR equation [Eqs.~(\ref{eq:SCReq2}) and (\ref{eq:SCReq3})] with respect to the scaled temperature $t$ under a constant volume: 
\begin{eqnarray}
\left(\frac{\partial{y}}{\partial{t}}\right)_{V}
=
\left\{
\begin{array}{lr}
\frac{\frac{y_1}{2t}\tilde{C}_{\rm b}\frac{1}{N}}{1-\frac{dy_1}{2t}M} 
\quad \mbox{for $d+z>4$},  \\
\frac{\frac{y_1}{2t}\tilde{C}_{\rm b}\frac{1}{N}}{1-\frac{y_1}{2}(\ln{y}+1)-\frac{dy_1}{2t}M}  
\quad \mbox{for $d+z=4$},
\end{array}
\right.
\label{eq:dydt}
\end{eqnarray}
where $M$ is given by
\begin{eqnarray}
M=
\int_{0}^{x_{\rm c}}dxx^{d+2z-5}
\left\{
\frac{1}{u}+\frac{1}{2u^2}-\psi'(u)
\right\}. 
\label{eq:M}
\end{eqnarray}
The low--$t$ behavior of $C_{\rm a}$ and $\tilde{C}_{\rm b}$ at the QCP for each class 
is summarized in the second and third columns of Table~\ref{tb:CaCb}~\cite{Makoshi1975,IM1996,MT,HM1995,Takahashi1999}, respectively. 
For each class, $C_{\rm a}$ dominates over $C_{\rm b}$ as $t$ decreases and hence 
the specific heat behaves as 
\begin{eqnarray}
C_{V}\approx C_{\rm a}
\end{eqnarray}
for $t\ll 1$. Hence, the criticality of $C_{V}$ is the same as that of $C_{\rm a}$ 
(see the last column of Table~\ref{tb:CaCb}). 
The criticality of $C_{V}$ in each class coincides with the RG theory~\cite{ZM1995,Zhu2003}.  

\begin{table}
\begin{center}
\begin{tabular}{l|cccc} \hline
{class} & {$y$} &{$C_{\rm a}$} & {$\tilde{C}_{\rm b}$}  & {$C_V$}
\\ \hline
3d FM &{$t^{\frac{4}{3}}$}  &{$-t\ln{t}$} & {$t^{\frac{4}{3}}$} & {$-t\ln{t}$}
\\ 
3d AFM &{$t^{\frac{3}{2}}$} &{const.$-t^{1/2}$} & {$t^{\frac{3}{2}}$} & {const.$-t^{1/2}$}
\\
2d FM &{$-t\ln{t}$} &{$t^{\frac{2}{3}}$} & {$-t\ln{t}$} &{$t^{\frac{2}{3}}$}
\\  
2d AFM &{$-\frac{t\ln{(-\ln{t})}}{\ln{t}}$} &
{$-t\ln{t}$}
& $t\ln(-\ln{t})$ & {$-t\ln{t}$}
\\ \hline
\end{tabular}
\end{center}
\caption{
Quantum criticality at the magnetic QCP for each class specified by $z=3$ (FM) and $z=2$ (AFM) in spatial dimension $d=3$ and $2$~\cite{Moriya,MU2003}. 
Temperature dependences of $C_{\rm a}$, $\tilde{C}_{\rm b}$, and $C_V$ at the QCP for $t\ll 1$~\cite{Makoshi1975,IM1996,MT,HM1995,Takahashi1999}.}
\label{tb:CaCb}
\end{table}

\section{Thermal Expansion Coefficient near 
the
 Magnetic QCP}
\label{sec:alpha}


Thus far, the thermal expansion coefficient $\alpha$ near magnetic transitions has been discussed with the spin-fluctuation theory~\cite{Takahashi2006} on the basis of the expression 
\begin{eqnarray}
\alpha
\equiv\frac{1}{V}\left(\frac{\partial V}{\partial T}\right)_{P}
=\kappa_{T}\left(\frac{\partial{P}}{\partial{T}}\right)_{V},
\label{eq:a_PT_def}
\end{eqnarray} 
where $\kappa_{T}$ is the isothermal compressibility defined as   
\begin{eqnarray}
\kappa_{T}\equiv-\frac{1}{V}\left(\frac{\partial{V}}{\partial{P}}\right)_{T}.
\label{eq:comp}
\end{eqnarray}
In this paper, we show that $\alpha$ can be expressed in a much simpler form, which enables us to capture the physical meaning. 
We start from the expression 
\begin{eqnarray}
\alpha=-\frac{1}{V}\left(\frac{\partial{S}}{\partial{P}}\right)_{T}  
\label{eq:a_SP_def}
\end{eqnarray}
equivalent to Eq.~(\ref{eq:a_PT_def}), which is transformed via the Maxwell relation $(\partial V/\partial T)_{P}=-(\partial S/\partial P)_{T}$. 
By differentiating the SCR entropy $S$ given by Eq.~(\ref{eq:S}) with respect to the pressure under a constant temperature, we obtain  
\begin{eqnarray} 
\left(\frac{\partial{S}}{\partial{P}}\right)_{T}
=
-\frac{C_{\rm a}}{T_0}\left(\frac{\partial{T_0}}{\partial{P}}\right)_{T}
-\frac{\tilde{C}_{\rm b}}{t}\left(\frac{\partial{y}}{\partial{P}}\right)_{T}, 
\end{eqnarray}
where $C_{\rm a}$ and $\tilde{C}_{\rm b}$ are given in Eqs.~(\ref{eq:Ca}) and (\ref{eq:tildaCb}), respectively. 
Here, we assume that the cutoff $x_{\rm c}$ in the $x$ integral in Eq.~(\ref{eq:S}) has no pressure dependence (and hence no volume dependence) since it has been supposed that the choice of the cutoff does not affect the low--energy physics in the SCR theory~\cite{Moriya}.
Then, we obtain the thermal expansion coefficient as 
\begin{eqnarray}
\alpha
=\alpha_{\rm a}+\alpha_{\rm b}, 
\label{eq:a_SP}
\end{eqnarray}
where $\alpha_{\rm a}$ and $\alpha_{\rm b}$ are defined as  
\begin{eqnarray}
\alpha_{\rm a}
&\equiv&
\frac{1}{V}
\frac{C_{\rm a}}{T_0}\left(\frac{\partial{T_0}}{\partial{P}}\right)_{T},
\label{eq:a_a}
\\
\alpha_{\rm b}
&\equiv&\frac{1}{V}
\frac{\tilde{C}_{\rm b}}{t}\left(\frac{\partial{y}}{\partial{P}}\right)_{T}, 
\label{eq:a_b}
\end{eqnarray}
respectively. 
Here, $(\partial{y}/\partial{P})_T$ can be calculated by differentiating the SCR equation 
[Eqs.~(\ref{eq:SCReq2}) and (\ref{eq:SCReq3})] with respect to the pressure under a constant temperature as 
\begin{eqnarray}
\left(\frac{\partial{y}}{\partial{P}}\right)_{T}
=
\left\{
\begin{array}{lr}
\frac{
\left(\frac{\partial{y_0}}{\partial{P}}\right)_{T}
+\left(\frac{\partial{y_1}}{\partial{P}}\right)_{T}
\frac{d}{2}
L
-\frac{1}{T_0}\left(\frac{\partial{T_0}}{\partial{P}}\right)_{T}
\tilde{C}_{\rm b}
\frac{y_1}{2}\frac{1}{N}
}
{
1-\frac{dy_1}{2t}M
}
\quad \ \mbox{$(d+z>4)$},  \\
\frac{
\left(\frac{\partial{y_0}}{\partial{P}}\right)_{T}
+\left(\frac{\partial{y_1}}{\partial{P}}\right)_{T}
\left(
\frac{d}{2}
L
+\frac{1}{2}y\ln{y}
\right)
-\frac{1}{T_0}\left(\frac{\partial{T_0}}{\partial{P}}\right)_{T}
\tilde{C}_{\rm b}
\frac{y_1}{2}\frac{1}{N}
}
{
1-\frac{y_1}{2}(\ln{y}+1)
-\frac{dy_1}{2t}M
}
\quad \ \mbox{$(d+z=4)$}, 
\end{array}
\right. 
\label{eq:dydP}
\end{eqnarray}
where $L$ is defined as  
\begin{eqnarray}
L\equiv
\int_{0}^{x_{\rm c}}dxx^{d+z-3}
\left\{
{\ln}u-\frac{1}{2u}-\psi(u)
\right\}.
\label{eq:L_def}
\end{eqnarray}
Equation~(\ref{eq:a_SP}) is one of the central results of this paper. 

The procedure for calculating $\alpha(t)$ is as follows:
First, we solve the SCR equation [Eq.~(\ref{eq:SCReq2}) or (\ref{eq:SCReq3})] by inputting $y_0=0$, which corresponds to the QCP, with setting $y_1=1$ and $x_{\rm c}=1$.  
Then, by using this solution $y(t)$, $C_{\rm a}(t)$ and $\tilde{C}_{\rm b}(t)$ are calculated as in Eqs.~(\ref{eq:Ca}) and (\ref{eq:tildaCb}), respectively. 
By inputting the solution $y(t)$ into Eq.~(\ref{eq:dydP}) with setting $(\partial y_0/\partial P)_{T}=1$, $(\partial y_1/\partial P)_{T}=1$, and $(\partial T_0/\partial P)_{T}/T_0=1$ as representative values (the reason for this parameterization is explained below), $(\partial y/\partial P)_T$ is obtained by calculating the right--hand side of Eq.~(\ref{eq:dydP}).
Finally, we obtain $\alpha_{\rm a}(t)$ in Eq.~(\ref{eq:a_a}) and $\alpha_{\rm b}(t)$ in Eq.~(\ref{eq:a_b}), resulting in $\alpha(t)$ in Eq.~(\ref{eq:a_SP}).

\begin{figure*}
\includegraphics[width=15cm]{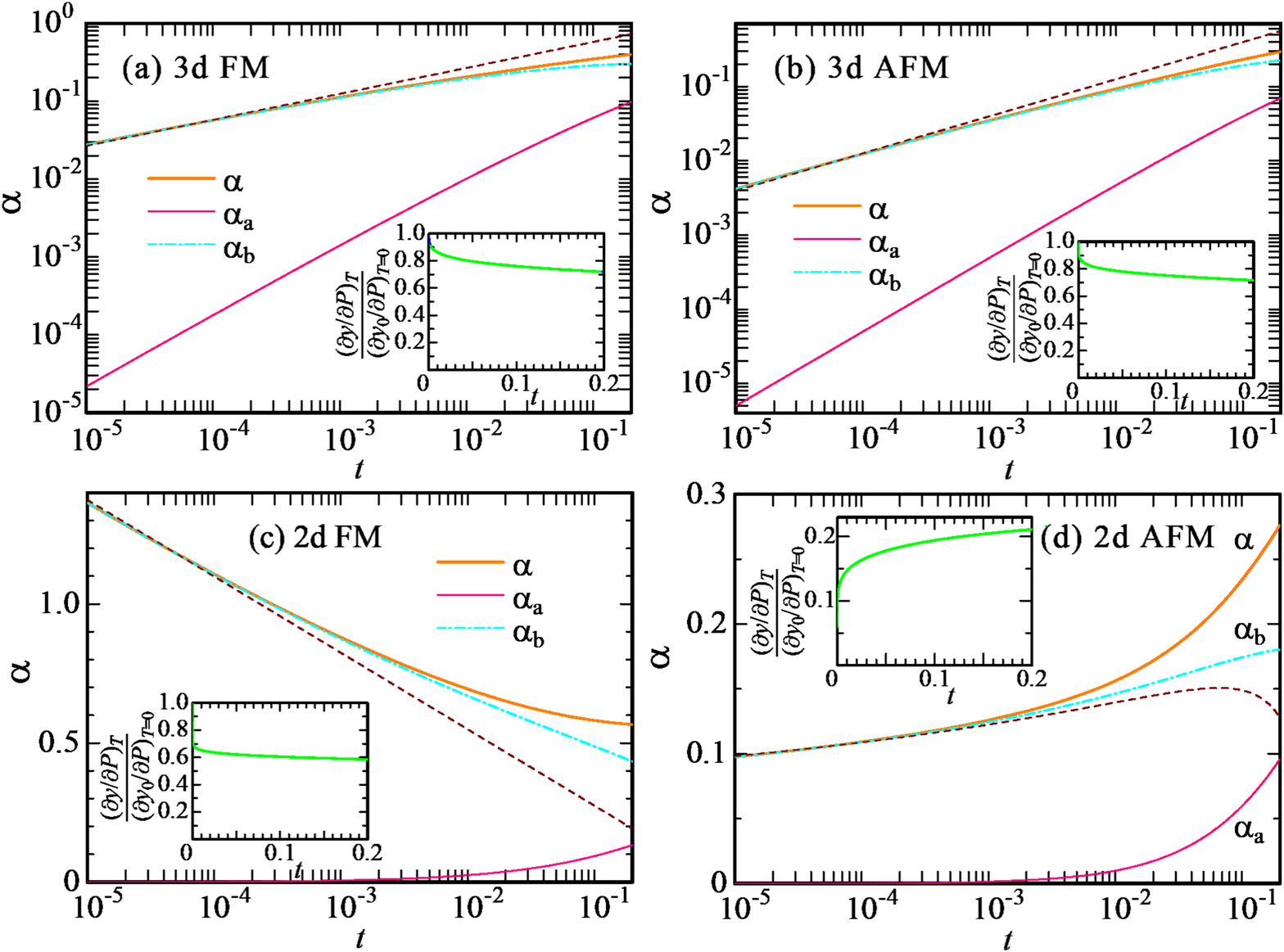}
\caption{(Color online) Thermal expansion coefficient $\alpha$ vs scaled temperature $t$ just at the QCP. 
The thick solid line is for $\alpha$, the thin solid line is for $\alpha_{\rm a}$, and the dashed-dotted line is for $\alpha_{\rm b}$.  
The dashed line represents the least-squares fit of $\alpha$ with $f_{\rm i}(t)$ ({\rm i}=a, b, c) for $10^{-5}\le t\le 10^{-4}$ and $f_{\rm d}(t)$ for $10^{-7}\le t\le 10^{-4}$. 
(a) 3d FM QCP: $f_{\rm a}(t)=at^{1/3}$.
(b) 3d AFM QCP: $f_{\rm b}(t)=at^{1/2}$.
(c) 2d FM QCP: $f_{\rm c}(t)=a\ln{t}$.
(d) 2d AFM QCP: $f_{\rm d}(t)=-a\frac{\ln(-\ln{t})}{\ln{\left(-\frac{t}{\ln{t}}\right)}}$. 
The inset shows the $t$ dependence of $\left(\frac{\partial{y}}{\partial{P}}\right)_{T}/\left(\frac{\partial{y_0}}{\partial{P}}\right)_{T=0}$. 
}
\label{fig:a_t_all}
\end{figure*}

Here, we note the unit and the parametrization of $\alpha$. When we input the value of $(\partial T_0/\partial P)_{T}/T_0$ in the unit of GPa$^{-1}$ and the molar volume as $V$ in the unit of \AA$^3$ into Eq.~(\ref{eq:a_a}), $\alpha_{\rm a}$ can be expressed in the unit of K$^{-1}$ as  
\begin{eqnarray}
\alpha_{\rm a}=\underline{\frac{0.0138}{V}\frac{1}{T_0}\left(\frac{\partial T_0}{\partial P}\right)_{T}}\times\frac{C_{\rm a}}{N}.
\label{eq:aa_Kinv}
\end{eqnarray}
When we input the value of $(\partial y/\partial P)_T$ in the unit of GPa$^{-1}$ and $V$ in the unit of \AA$^3$ into Eq.~(\ref{eq:a_b}), $\alpha_{\rm b}$ can be expressed in the unit of K$^{-1}$ as
\begin{eqnarray}
\alpha_{\rm b}=\underline{\frac{0.0138}{V}\left(\frac{\partial y}{\partial P}\right)_{T}}\times\frac{\tilde{C}_{\rm b}}{Nt}.
\label{eq:ab_Kinv} 
\end{eqnarray}
Hence, multiplying the numerical values of the underlined terms in Eqs.~(\ref{eq:aa_Kinv}) and (\ref{eq:ab_Kinv}) for each material to the following results of $\alpha_{\rm a}$ and $\alpha_{\rm b}$, respectively, shown in Figs.~\ref{fig:a_t_all}(a)--\ref{fig:a_t_all}(d), one can make a direct comparison with experiments.

The results of the numerical calculations of $\alpha(t)$ for the input parameters mentioned above for each universality class are shown in Figs.~\ref{fig:a_t_all}(a)--\ref{fig:a_t_all}(d). 
Note that, in each class, $\alpha_{\rm b}$ becomes dominant as $t$ decreases, while $\alpha_{\rm a}$ gives a minor contribution.  

For $d+z>4$, $\alpha$ behaves as
\begin{eqnarray}
\alpha\propto\frac{\tilde{C}_{\rm b}}{t}
\label{eq:tldCbt}
\end{eqnarray}
for $t\ll 1$, where $(\partial y/\partial P)_{T}$ is regarded as temperature independent, which is verified by the almost $t$--independent behavior shown in the inset of Figs.~\ref{fig:a_t_all}(a)--\ref{fig:a_t_all}(c). 
The critical behavior expressed as Eq.~(\ref{eq:tldCbt}) actually appears for $t\ll 1$ as shown by the dashed line in Figs.~\ref{fig:a_t_all}(a)--\ref{fig:a_t_all}(c), which is summarized in the first column of Table~\ref{tb:alpha_QCP}. 
The thermal expansion coefficient in the 2d FM case diverges $\alpha(t)\to\infty$ for $t\to 0$, whereas $\alpha(t)\to 0$ for $t\to 0$ in the 3d FM and 3d AFM cases. 
These $t$ dependences coincide with those shown by the RG theory~\cite{Zhu2003}. 
However, this asymptotic criticality appears only at sufficiently low temperatures for $t\lsim 10^{-3}$ in the 3d FM case [Fig.~\ref{fig:a_t_all}(a)] and the 3d AFM case [Fig.~\ref{fig:a_t_all}(b)], and for $t\lsim 10^{-4}$ in the 2d FM case [Fig.~\ref{fig:a_t_all}(c)]. 
This is due to the presence of the weakly temperature-dependent $(\partial y/\partial P)_{T}$ in $\alpha_{\rm b}$, as shown in the inset of Figs.~\ref{fig:a_t_all}(a)--\ref{fig:a_t_all}(c). 

For the marginal case $d+z=4$, i.e., the 2d AFM case, $\alpha(t)$ for $t\ll 1$ is evaluated as 
\begin{eqnarray}
\alpha
\propto\frac{\tilde{C}_{\rm b}}{t}\left(\frac{\partial{y}}{\partial{P}}\right)_{T}
\sim-\frac{\ln(-\ln{t})}{\ln\left(-\frac{t}{\ln{t}}\right)}.
\label{eq:a_t_d2z2}
\end{eqnarray}
This can be seen in Fig.~\ref{fig:a_t_all}(d), where $\alpha$ is well fit by the dashed line expressed as  Eq.~(\ref{eq:a_t_d2z2}).  
Here, we found that $(\partial y/\partial P)_{T}$ has the temperature dependence even for $t\ll 1$ as $(\partial y/\partial P)_{T}\approx (\partial y_0/\partial P)_{T=0}(-b_4)/\ln(-t/\ln t)$ with $b_4$ being a positive constant, which can be confirmed in the inset of Fig.~\ref{fig:a_t_all}(d). 
This is due to the logarithmic correction term in Eq.~(\ref{eq:SCReq3}). 
Namely, $(\partial y/\partial P)_{T}$ affects the criticality in Eq.~(\ref{eq:a_t_d2z2}), which was not reported in the past RG studies~\cite{Zhu2003,Garst2005}.
Hence, the $t$ dependence of $\tilde{C}_{\rm b}/t$ showing divergence as $\sim\ln(-\ln{t})$ for $t\to 0$  (see the third column in Table~\ref{tb:CaCb}), in agreement with the RG theory~\cite{Zhu2003}, is counteracted by $(\partial y/\partial P)_{T}\to 0$ for $t\to 0$.

\begin{table}
\begin{center}
\begin{tabular}{l|cc} \hline
{class} &{$\alpha$} & {$\Gamma$}  
\\ \hline
3d FM &{$t^{1/3}$} & {$-\frac{t^{-2/3}}{\ln{t}}$} 
\\ 
3d AFM &{$t^{1/2}$} & {$\frac{t^{-1/2}}{{\rm const.}-t^{1/2}}$}  
\\
2d FM &{$-\ln{t}$} & {$-t^{-2/3}\ln{t}$} 
\\  
2d AFM &
{$-\frac{\ln{(-\ln{t})}}{\ln\left(-\frac{t}{\ln{t}}\right)}$}
& $\frac{1}{t\ln{t}}\frac{\ln(-\ln{t})}{\ln\left(-\frac{t}{\ln{t}}\right)}$ 
\\ \hline
\end{tabular}
\end{center}
\caption{Temperature dependences of $\alpha$ and $\Gamma$ at the QCP of FM $(z=3)$ and AFM $(z=2)$
in spatial dimensions $(d=3,2)$.}
\label{tb:alpha_QCP}
\end{table}

\section{Gr\"{u}neisen Parameter near the Magnetic QCP}

The Gr\"{u}neisen parameter $\Gamma$ is defined by
\begin{eqnarray}
\Gamma\equiv\frac{\alpha{V}}{C_{V}\kappa_{T}}. 
\label{eq:Grn}
\end{eqnarray}
Since $\alpha$ is expressed as $\alpha_{\rm a}+\alpha_{\rm b}$, $\Gamma$ [Eq.~(\ref{eq:Grn})] is expressed as
\begin{eqnarray}
\Gamma=\Gamma_{\rm a}+\Gamma_{\rm b},
\label{eq:Grn_ab}
\end{eqnarray}
where $\Gamma_{\rm i}$ $({\rm i}={\rm a}, {\rm b})$ is defined by 
\begin{eqnarray}
\Gamma_{\rm i}\equiv\frac{\alpha_{\rm i}{V}}{C_{V}\kappa_{T}}. 
\label{eq:Grn_i}
\end{eqnarray}

At low temperatures, $C_{V}$ is governed by $C_{\rm a}$ as $C_{V}=C_{\rm a}-C_{\rm b}\approx C_{\rm a}$. 
By using Eqs.~(\ref{eq:a_a}) and (\ref{eq:a_b}), $\Gamma$ [Eq.~(\ref{eq:Grn_ab})] is expressed for $t\ll 1$ as 
\begin{eqnarray}
\Gamma&\approx&\frac{1}{\kappa_T}\frac{1}{T_0}\left(\frac{\partial{T_0}}{\partial{P}}\right)_{T}
+\frac{\tilde{C}_{\rm b}}{C_{\rm a}}\frac{1}{t}\frac{1}{\kappa_{T}}\left(\frac{\partial{y}}{\partial{P}}\right)_{T},
\nonumber
\\
&=&
-\frac{V}{T_0}\left(\frac{\partial{T_0}}{\partial{V}}\right)_{T}
-\frac{\tilde{C}_{\rm b}}{C_{\rm a}}\frac{V}{t}\left(\frac{\partial{y}}{\partial{V}}\right)_{T}, 
\label{eq:Grn_lowt}
\end{eqnarray} 
where Eq.~(\ref{eq:comp}) has been used to derive the second line. 
One can see that the first and second terms of Eq.~(\ref{eq:Grn_lowt}) correspond to $\Gamma_{\rm a}$
 and $\Gamma_{\rm b}$, respectively. 

Here, we note that Eq.~(\ref{eq:Grn_lowt}) is consistent with  
$\Gamma$ derived under an adiabatic process. 
By differentiating both sides of Eq.~(\ref{eq:S}) with respect to the volume under a constant entropy
and using the expression $\Gamma=-\frac{V}{T}\left(\frac{\partial{T}}{\partial{V}}\right)_{S}$, which is  equivalent to Eq.~(\ref{eq:Grn}), we obtain 
\begin{eqnarray}
\Gamma=-\frac{V}{T_0}
\left(\frac{\partial{T_0}}{\partial{V}}\right)_{S}
-\frac{\tilde{C}_{\rm b}}{C_{\rm a}}\frac{V}{t}\left(\frac{\partial{y}}{\partial{V}}\right)_{S}. 
\label{eq:Grn_QCP1}
\end{eqnarray}
We see that the first and second terms correspond to those in Eq.~(\ref{eq:Grn_lowt}), respectively. 
 
As for the first term in Eq.~(\ref{eq:Grn_lowt}), $\Gamma_{\rm a}(T=0)=-\frac{V}{T_0}\left(\frac{\partial{T_0}}{\partial{V}}\right)_{T=0}$ is the volume derivative of the characteristic temperature of spin fluctuation. 
In heavy electron systems, $|\Gamma_{\rm a}|$ typically has an enhanced value with $O(10)$ being in the same order of the Gr\"{u}neisen parameter in the Fermi-liquid regime $\Gamma_{\rm FL}$. Here, $\Gamma_{\rm FL}$ is defined as  
$\Gamma_{\rm FL}\equiv-\frac{V}{T_{\rm K}}\left(\frac{\partial{T_{\rm K}}}{\partial{V}}\right)_{T=0}$, where $T_{\rm K}$ is the characteristic temperature of heavy electrons called Kondo temperature.
Since $\Gamma_{\rm FL}$ typically has an enhanced value~\cite{Flouquet2005,Goltsev2005,UmeoPRB1996} of $O(10)$ and $T_0$ is shown to be proportional to $T_{\rm K}$, $|\Gamma_{\rm a}|$ is also enhanced 
although $|\Gamma_{\rm a}(t)|$ is almost $t$ independent.  

To analyze the $t$ dependence of $\Gamma$ at the QCP, we performed the numerical calculation of Eq.~(\ref{eq:Grn}). 
We calculate $\alpha$ by the procedure in Sect.~\ref{sec:alpha}. 
As for $C_{V}$, we calculate $C_{\rm a}$ and $C_{\rm b}$ in Eqs.~(\ref{eq:Ca}) and (\ref{eq:Cb}), respectively, where $(\partial y/\partial t)_V$ is obtained by calculating Eq.~(\ref{eq:dydt}). 
The input parameters set is the same as that set for Fig.~\ref{fig:a_t_all}. 
As for the isothermal compressibility, we confirmed that $\kappa_{T}$ does not show divergence even at the QCP for each class but has a finite value in general and hence we input $\kappa_{T}=0.1$ as a typical value for heavy electron systems. 
This is because $\frac{1}{T_0}\left(\frac{\partial{T_0}}{\partial{P}}\right)_{T}=\Gamma_{\rm a}(T=0)\kappa_{T}=1$ was used in Sect.~\ref{sec:alpha} and $\kappa_{T}$ is set so as to reproduce $\Gamma_{\rm a}(T=0)=10$. 

\begin{figure*}
\includegraphics[width=15cm]{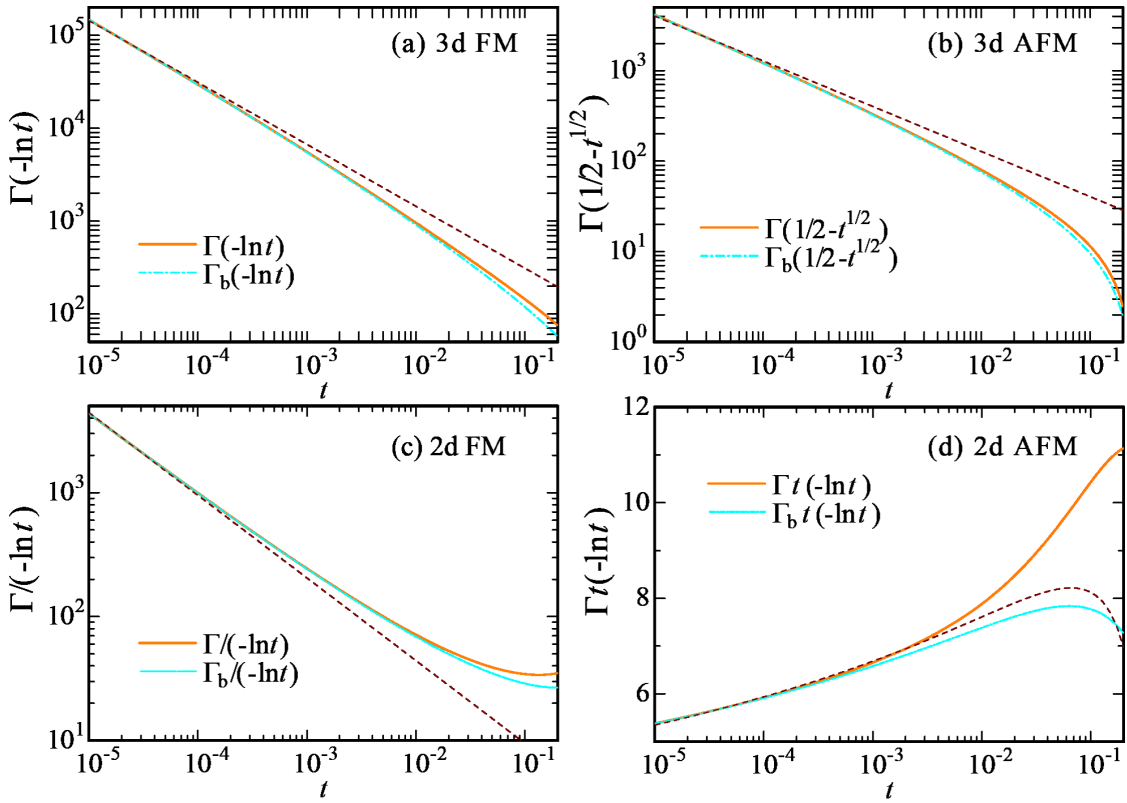}
\caption{(Color online) Temperature dependence of Gr\"{u}neisen parameter 
$\Gamma$ (thick solid line) and $\Gamma_{\rm b}$ (dash-dotted line) just at the QCP.  
The dashed line represents the least-square fit of $\alpha$ with $g_{\rm i}(t)$ ({\rm i}=a, b, c, d) for $10^{-5}\le t\le 10^{-4}$. 
(a) 3d FM QCP: $g_{\rm a}(t)=at^{-2/3}$.
(b) 3d AFM QCP: $g_{\rm b}(t)=at^{-1/2}$.
(c) 2d FM QCP: $g_{\rm c}(t)=at^{-2/3}$.
(d) 2d AFM QCP: $g_{\rm d}(t)=-a\frac{\ln(-\ln{t})}{\ln\left(-\frac{t}{\ln{t}}\right)}$. 
}
\label{fig:G_t_all}
\end{figure*}

The results of the numerical calculations for each class are shown in Figs.~\ref{fig:G_t_all}(a)--\ref{fig:G_t_all}(d). 
Since $\Gamma_{\rm a}$ has a minor $t$ dependence here, we show the $t$ dependences of $\Gamma$ and $\Gamma_{\rm b}$. 
Reflecting the fact that $\alpha$ is dominated by $\alpha_{\rm b}$ for low $t$ (see Fig.~\ref{fig:a_t_all}), $\Gamma$ is mainly contributed from $\Gamma_{\rm b}$. 
Then, as $t$ decreases, $\Gamma$ increases and finally diverges for $t\to 0$ in each class because of the factor $1/t$ in the last term of Eq.~(\ref{eq:Grn_lowt}). 
Our analysis has revealed 
that the divergence of the Gr\"{u}neisen parameter arises from the term with the volume derivative of  the inverse susceptibility of spin fluctuation [see the last term of Eq.~(\ref{eq:Grn_lowt})].

For $d+z>4$, $\Gamma$ behaves as 
\begin{eqnarray}
\Gamma\propto -\frac{\tilde{C}_{\rm b}}{C_{\rm a}}\frac{1}{t}
\end{eqnarray}
for $t\ll 1$ because $(\partial y/\partial P)_{T}$ is almost $t$ independent as mentioned in Sect.~3. 
The critical $t$ dependence for each class, which can be known analytically from Table~\ref{tb:CaCb}, is actually confirmed by the numerical result well fitted by the dashed line in Figs.~\ref{fig:G_t_all}(a)--\ref{fig:G_t_all}(c). 
The criticality of $\Gamma$ is summarized in the second column of Table~\ref{tb:alpha_QCP}, which is in agreement with the RG result~\cite{Zhu2003}. 
Note, however, that the criticality appears at a sufficiently low--$t$ regime for $t\lsim 10^{-3}$, indicating that the temperature dependence of $(\partial y/\partial P)_{T}$ affects the intermediate--$t$ region.  

For $d+z=4$, i.e., the 2d AFM case, $\Gamma$ behaves as
\begin{eqnarray}
\Gamma\propto -\frac{\tilde{C}_{\rm b}}{C_{\rm a}}\frac{V}{t}\left(\frac{\partial{y}}{\partial{V}}\right)_{T}
\sim\frac{1}{t\ln{t}}\frac{\ln(-\ln{t})}{\ln\left(-\frac{t}{\ln{t}}\right)}
\end{eqnarray}
for $t\ll 1$. 
The critical $t$ dependence comes from the multiplication of $\tilde{C}_{\rm b}/(C_{\rm a}t)$ (see Table~\ref{tb:CaCb}) and the prefactor $(\partial y/\partial P)_{T}\approx 1/\ln(-t/\ln{t})$, which can be confirmed numerically by the dashed line in Fig.~\ref{fig:G_t_all}(d). 
The $t$ dependence of $\tilde{C}_{\rm b}/(C_{\rm a}t)$ for $t\ll 1$ is in agreement with the RG theory~\cite{Zhu2003}. Even after the inclusion of the $t$ dependence of $(\partial y/\partial P)_{T}$, $\Gamma$ diverges for $t\to 0$ since the factor $1/t$ in $\tilde{C}_{\rm b}/(C_{\rm a}t)$ overcomes the logarithmic correction. 

\section{Discussion}

To observe $\alpha(T)$ and $\Gamma(T)$ near the magnetic QCP, experimental measurements have been performed~\cite{Kambe1997,Kuchler2003,Kuchler2006,Kuchler2007,Steppke2013,Gegenwart2016}. So far, a few data have been reported in stoichiometric compounds, which follow the criticality shown in Tables~\ref{tb:CaCb} and \ref{tb:alpha_QCP}.

CeNi$_2$Ge$_2$ at ambient pressure is regarded to be located closely to the 3d AFM QCP since the low--$T$ data of the specific heat and resistivity show the 3d--AFM criticality in Table~\ref{tb:CaCb}~\cite{Kuchler2003}. 
The measured thermal expansion coefficient $\alpha=c_{1}\sqrt{T}+c_{2}T$ is in agreement with 
$\alpha\sim T^{1/2}$ in Table~\ref{tb:CaCb} induced by spin fluctuation arising from the QCP 
and the FL contribution $\alpha_{\rm FL}\sim T$. 
The Gr\"{u}neisen parameter $\Gamma\approx 57$ at $T=5$~K is already enhanced because of the contribution from $\Gamma_{\rm a}$ and the heavy--electron background $\Gamma_{\rm FL}$. 
As $T$ decreases, $\Gamma$ further increases as $\Gamma\approx 98\pm 10$ at $T\approx 0.1$~K, 
indicating the contribution from $\Gamma_{\rm b}$.

In Ce$_7$Ni$_3$, the 3d AFM ordering is suppressed by applying pressure $P$ around $P_{\rm c}=0.39$~GPa. 
As pressure increases, $T_{\rm K}(P)$ and $T_{0}(P)$ increase, and a smooth variation of both in the $T$-$P$ phase diagram was observed~\cite{Umeo1996,UmeoPRB1996}. 
This is understandable from the relation $T_0\propto T_{\rm K}$ as noted below Eq.~(\ref{eq:Grn_QCP1}). 
The measurements of $\alpha(T)$ and $\Gamma(T)$ at the QCP and their analyses based on Eqs.~(\ref{eq:a_SP}) and (\ref{eq:Grn_ab}) are interesting studies left for the future.

\section{Summary}

On the basis of the SCR theory of spin fluctuations, we have derived the analytical expressions of the thermal expansion coefficient and the Gr\"{u}neisen parameter near the magnetic QCP 
and have numerically analyzed their properties. 

The specific heat under a constant volume is expressed as
$C_{V}=C_{\rm a}-C_{\rm b}$, where $C_{\rm b}$ has the form $C_{\rm b}=\tilde{C}_{\rm b}\left(\frac{\partial y}{\partial t}\right)_{V}$. 
We have derived the explicit forms of $\left(\frac{\partial y}{\partial t}\right)_{V}$ from the SCR equations for $d+z>4$ and $d+z=4$, respectively. 

We have derived the thermal expansion coefficient from the expression of the SCR entropy through the relation $\alpha=-(\partial S/\partial P)_{T}/V$ as 
$\alpha=\alpha_{\rm a}+\alpha_{\rm b}$ for each class, where $\alpha_{\rm a}=\frac{1}{V}
\frac{C_{\rm a}}{T_0}\left(\frac{\partial{T_0}}{\partial{P}}\right)_{T}$
and
$\alpha_{\rm b}=\frac{\tilde{C}_{\rm b}}{t}\left(\frac{\partial{y}}{\partial{P}}\right)_{T}$.
We have found that at low temperatures, $\alpha_{\rm b}$ dominates over $\alpha_{\rm a}$,  
while $C_{\rm a}$ dominates over $C_{\rm b}$ in each class.  
An important result is that there exists a temperature--dependent prefactor $(\partial y/\partial P)_{T}$ in $\alpha_{\rm b}$, which contributes to the intermediate--temperature region between the Curie--Weiss regime and quantum--critical regime. 
Furthermore, $(\partial y/\partial P)_{T}$ even affects the quantum criticality for $d+z=4$, i.e., the 2d AFM case, giving rise to $\alpha(T)\to 0$ as $-\ln(-\ln t)/\ln(-t/\ln t)$, with $t=T/T_{0}$, for $T\to 0$.  

On the basis of these correctly calculated $C_V$ and $\alpha$, we have derived the Gr\"{u}neisen parameter. We have shown that $\Gamma$ is expressed as  
$\Gamma=\Gamma_{\rm a}+\Gamma_{\rm b}$, where $\Gamma_{\rm i}=\frac{\alpha_{i}V}{C_{V}\kappa_{T}}$ $({\rm i}={\rm a}, {\rm b})$. 
At low temperatures, 
$\Gamma_{\rm a}$ shows a minor $T$ dependence and is expressed as 
$\Gamma_{\rm a}=-\frac{V}{T_0}\left(\frac{\partial T_0}{\partial V}\right)_T$ for $T\to 0$, which has an enhanced value of typically $O(10)$ in the heavy electron systems. 
As temperature decreases, $\Gamma$ further increases, which is mainly contributed from $\Gamma_{\rm b}\approx-\frac{\tilde{C}_{\rm b}}{C_{\rm a}}\frac{V}{t}\left(\frac{\partial{y}}{\partial{V}}\right)_{T}$, and $\Gamma$ finally diverges for $T\to 0$ in each class. 
Our analysis has revealed that the divergence of the Gr\"{u}neisen parameter arises from the inverse susceptibility of spin fluctuations coupled to the volume. 

Numerical calculations of $\alpha(T)$ and $\Gamma(T)$ for each class show that the quantum--critical temperature dependence appears in the sufficiently low $T$ regime, which is typically below $T/T_0\lsim 10^{-3}$ with $T_0$ being the characteristic temperature of spin fluctuation, owing to the $T$--dependent prefactor $(\partial y/\partial P)_{T}$. 
This is important when one makes a comparison with experiments.

\section*{Acknowledgment}

The authors are grateful to K. Umeo for discussions about experimental data of Ce$_7$Ni$_3$.
This work was supported by Grants-in-Aid for Scientific Research (Grant Numbers JP24540378, JP25400369, 
JP15K05177, JP16H01077, and JP17K05555).


\begin{thebibliography}{99}
\bibitem{Moriya} T. Moriya and A. Kawabata, J. Phys. Soc. Jpn. {\bf 34}, 639 (1973); T. Moriya, {\it Spin Fluctuations in Itinerant Electron Magnetism} (Springer-Verlag, Berlin, 1985). 
\bibitem{HM1995} M. Hatatani and T. Moriya, J. Phys. Soc. Jpn. {\bf 64}, 3434 (1995). 
\bibitem{MT} T. Moriya and T. Takimoto, J. Phys. Soc. Jpn. {\bf 64}, 960 (1995). 
\bibitem{IM1996} A. Ishigaki and T. Moriya, J. Phys. Soc. Jpn. {\bf 65}, 376 (1996). 
\bibitem{IM1998} A. Ishigaki and T. Moriya, J. Phys. Soc. Jpn. {\bf 67}, 3924 (1998). 
\bibitem{Ishigaki} A. Ishigaki and T. Moriya, J. Phys. Soc. Jpn. {\bf 68}, 3673 (1999). 
\bibitem{Hertz} J. A. Hertz, Phys. Rev. B {\bf 14}, 1165 (1976).
\bibitem{Millis} A. J. Millis, Phys. Rev. B {\bf 48}, 7183 (1993);  
the statement ``SCR procedure does not yield the log corrections in $d=2$ for $z=3$ and $z=2$." is not correct. The SCR results coincide with those derived by the renormalization group theory including the log corrections in $d=2$ for $z=3$~\cite{HM1995} and $z=2$ (T. Moriya: private communications). 
\bibitem{MU1980} T. Moriya and K. Usami, Solid State Commun. {\bf 34}, 95 (1980). 
\bibitem{Gruneisen1912} E. Gr{\"u}neisen, Ann. Phys. {\bf 39}, 257 (1912).
\bibitem{Kambe1997} S. Kambe, J. Flouquet, P. Lejay, P. Haen, and A. de Visser, J. Phys.: Condens. Matter {\bf 9}, 4917 (1997). 
\bibitem{Zhu2003} L. Zhu, M. Garst, A. Rosch, and Q. Si, Phys. Rev. Lett. {\bf 91}, 066404 (2003). 
\bibitem{Garst2005} M. Garst and A. Rosch, Phys. Rev. B {\bf 72}, 205129 (2005). 
\bibitem{Takahashi2006} Y. Takahashi and H. Nakano, J. Phys.: Condens. Matter {\bf 18}, 521 (2006). 
\bibitem{note_Heisenberg} The present results can be extended straightforwardly to the case of magnetic fluctuations of XY or Heisenberg type, where the factor of 2 or 3 is multiplied to the entropy in Eq.~(\ref{eq:S}) and the specific heat in Eqs.~(\ref{eq:Ca}) and (\ref{eq:tildaCb}). In real materials, anisotropy of the spin space more or less exists, yielding anisotropic spin fluctuation. Therefore, we plot $\alpha$ for a single component of the spin fluctuation with the factor of 1 in Fig.~\ref{fig:a_t_all} for convenience of comparison with experiments. 
As for the Gr\"{u}neisen parameter $\Gamma$, the results shown in Fig.~\ref{fig:G_t_all} are not affected by this factor 
except for $\kappa_{T}$, since the factors in $\alpha$ and $C_{V}$ cancel each other in Eq.~(\ref{eq:Grn_i}). 
\bibitem{Takahashi1999} Y. Takahashi, J. Phys.: Condens. Matter {\bf 11}, 6439 (1999). 
\bibitem{Feynman} R. P. Feynman, {\it Statistical Mechanics} (Addison-Wesley, Reading, Massachusetts, 1990) Sect. 3.4.
\bibitem{MU2003} T. Moriya and K. Ueda, Rep. Prog. Phys. {\bf 66}, 1299 (2003).
\bibitem{Makoshi1975} K. Makoshi and T. Moriya, J. Phys. Soc. Jpn. {\bf 38}, 10 (1975).  
\bibitem{ZM1995} U. Z\"{u}licke and A. J. Millis, Phys. Rev. B {\bf 51}, 8996 (1995).  
\bibitem{UmeoPRB1996} K. Umeo, H. Kadomatsu, and T. Takabatake, Phys. Rev. B {\bf 54}, 1194 (1996). 
\bibitem{Flouquet2005} J. Flouquet, Prog. Low. Temp. Phys. {\bf 15}, 149 (2005). 
\bibitem{Goltsev2005} A. V. Goltsev and M. M. Abd-Elmeguid, J. Phys.: Condens. Matter {\bf 17}, 5813 (2005). 
\bibitem{Kuchler2003} R. K{\"u}chler, N. Oeschler, P. Gegenwart, T. Cichorek, K. Neumaier, O.~Tegus, C. Geibel, J. A. Mydosh, F. Steglich, L. Zhu, and Q. Si, Phys. Rev. Lett. {\bf 91}, 066405 (2003). 
\bibitem{Kuchler2006} R. K{\"u}chler, P. Gegenwart, J. Custers, O. Stockert, N. Caroca-Canales, C.~Geibel, J. G. Sereni, and F. Steglich, Phys. Rev. Lett. {\bf 96}, 256403 (2006). 
\bibitem{Kuchler2007} R. K\"{u}chler, P. Gegenwart, C. Geibel, and F. Steglich, Sci. Technol. Adv. Mater. {\bf 8}, 428 (2007) and references therein. 
\bibitem{Steppke2013} A. Steppke, R. Kuchler, S. Lausberg, E. Lengyel, L. Steinke, 
R. Borth, T. Luhmann, C. Krellner, M. Nicklas, C. Geibel, F. Steglich, and M.~Brando, Science {\bf 339}, 933 (2013). 
\bibitem{Gegenwart2016} P. Gegenwart, Rep. Prog. Phys. {\bf 79}, 114502 (2016) and references therein.
\bibitem{Umeo1996} K. Umeo, H. Kadomatsu, and T. Takabatake, J. Phys.: Condens. Matter {\bf 8}, 9743 (1996).  
\end{thebibliography}
\end{document}